\documentclass[fleqn,10pt]{wlscirep}
\DeclareGraphicsExtensions{.eps,.ps,.jpg,.bmp,.pdf}

\usepackage{multirow}
\usepackage{graphicx}
\usepackage{bm}
\makeatletter
\renewcommand\normalsize{
   \@setfontsize\normalsize\@xpt\@xiipt
   \abovedisplayskip 8\p@ \@plus8\p@ \@minus8\p@
   \abovedisplayshortskip \z@ \@plus8\p@
   \belowdisplayshortskip 8\p@ \@plus8\p@ \@minus8\p@
   \belowdisplayskip \abovedisplayskip
   \let\@listi\@listI}
\makeatother

\title{Adaptive pixel-super-resolved lensfree holography for wide-field on-chip microscopy}

\author[1,2,3]{Jialin Zhang}
\author[1,2,3,+]{Jiasong Sun}
\author[1,2,*]{Qian Chen}
\author[1,2,3]{Jiaji Li}
\author[1,2,3,*]{Chao Zuo}

\affil[1]{School of Electronic and Optical Engineering, Nanjing University of Science and Technology, No. 200 Xiaolingwei Street, Nanjing, Jiangsu Province 210094, China}
\affil[2]{Jiangsu Key Laboratory of Spectral Imaging \& Intelligent Sense, Nanjing, Jiangsu Province 210094, China}
\affil[3]{Smart Computational Imaging Laboratory (SCILab), Nanjing University of Science and Technology, Nanjing, Jiangsu Province 210094, China}

\affil[*]{Correspondence and requests for materials should be addressed to C.Z. (email: zuochao@njust.edu.cn) or Q.C. (email: chenqian@njust.edu.cn)}
\affil[+]{this author contributed equally to this work}


\begin{abstract}
High-resolution wide field-of-view (FOV) microscopic imaging plays an essential role in various fields of biomedicine, engineering, and physical sciences. As an alternative to conventional lens-based scanning techniques, lensfree holography provides a new way to effectively bypass the intrinsical trade-off between the spatial resolution and FOV of conventional microscopes. Unfortunately, due to the limited sensor pixel-size, unpredictable disturbance during image acquisition, and sub-optimum solution to the phase retrieval problem, typical lensfree microscopes only produce compromised imaging quality in terms of lateral resolution and signal-to-noise ratio (SNR). Here, we propose an adaptive pixel-super-resolved lensfree imaging (APLI) method which can solve, or at least partially alleviate these limitations. Our approach addresses the pixel aliasing problem by Z-scanning only, without resorting to subpixel shifting or beam-angle manipulation. Automatic positional error correction algorithm and adaptive relaxation strategy are introduced to enhance the robustness and SNR of reconstruction significantly. Based on APLI, we perform full-FOV reconstruction of a USAF resolution target ($\sim$29.85 $m{m^2}$) and achieve half-pitch lateral resolution of $770$ $nm$, surpassing $2.17$ times of the theoretical Nyquist–Shannon sampling resolution limit imposed by the sensor pixel-size ($1.67$ $\mu m$). Full-FOV imaging result of a typical dicot root is also provided to demonstrate its promising potential applications in biologic imaging.
\end{abstract}

\begin{document}

\flushbottom
\maketitle
\thispagestyle{empty}

\section*{Introduction}
High-resolution wide-field optical imaging is an essential tool in various biomedical applications \cite{maricq1973patterns,huisman2010creation} including cell cycle assay, digital pathology, and high-throughput biologic screening. The growing need in digitalizing the biological slides has facilitated the development of whole slide imaging (WSI) systems.  Nevertheless, these systems are built based on a conventional microscope, which suffer from the inherent trade-off between the field-of-view (FOV) and imaging resolution. To get an image with both high resolution and large FOV, mechanical scanning and stitching is required to expand the limited FOV of a conventional high magnification objective \cite{ma2007use}, which not only complicate the imaging procedure, but also significantly increase the overall cost of these systems. The recently developed computational microscopy techniques provide new opportunities to create high-resolution wide-field images without any scanning and stitching, such as synthetic aperture microscopy \cite{mico2006synthetic,hillman2009high,feng2009long,tippie2011high}, Fourier ptychography microscopy (FPM) \cite{zheng2013wide,tian2014multiplexed,sun2017resolution}, and lens-free super-resolution holography \cite{bishara2010lensfree,zheng2011epetri,greenbaum2014wide}. Among these approaches, the lens-free super-resolution holography has unique advantages of achieving a large effective numerical aperture (NA) approaching to unity across the native FOV of the imaging sensor, without requiring any lenses and other intermediate optical components. This further allows to significantly simplify the imaging setup and meanwhile effectively circumvent the optical aberrations and chromaticity \cite{zheng2011epetri} that are inherent in conventional lens-based imaging systems. Besides, the whole system can be built in a miniaturized and cost-effective format, providing a potential solution for reducing health care costs for point-of-care diagnostics in resource-limited environments.

In recent years, numerous lensfree image systems have been proposed, and most of them are unit magnification systems in which the samples are placed as close as possible to the the imaging devices. These unit magnification systems can not only reduce the demand on the coherence of the illumination source but also have a significantly large FOV claiming the entire digital sensor area as the microscopic imaging FOV compared to the traditional amplification ones \cite{haddad1992fourier,xu2001digital,pedrini2002short,repetto2004lensless,garcia2006immersion}. However, these lensfree holographic microscopes suffer from low resolution which is far away from the demand of recent high-resolution-imaging applications. According to Nyquist–Shannon sampling theorem \cite{shannon1949communication}, the resolution of the reconstruction results is limited to the sampling resolution of the imaging devices. In other words, the physical pixel-size will be the main limiting factor of the lensfree imaging systems, because the systems will fail to record the diffraction fringes carrying high spatial frequencies which will result in the aliasing of the spectrum. To improve the resolution of the reconstruction results, one way is to physically reduce the pixel-size that can directly achieve better spatial resolution for both digital in-line holographic and contact-mode lensfree microscopy, unless pixel design exhibits severe angular distortions creating aberrations for oblique rays \cite{greenbaum2012imaging}. Nevertheless, the physical reduction of pixel-size will sacrifice signal-to-noise ratio (SNR) due to the reduction of the external quantum efficiency on the smaller light sensing area \cite{chen2000small}. Moreover, the smaller pixel-size is a major development trend of the commercial sensor chips, but the pixel-size of the available sensor still cannot satisfy the rapidly growing demands in lensfree in-line holographic microscopes due to the obstacles of the technology of semiconductor manufacturing. Besides reducing physical pixel-size, another way is to digitally synthesize a much smaller effective pixel-size termed pixel super-resolution methods \cite{greenbaum2014wide,luo2015synthetic} which refer to the de-aliasing, rather than surpassing the diffraction limit of light. In the past few years, many pixel super-resolution methods have been proposed by shifting the illumination source, the imaging devices or the samples \cite{bishara2010lensfree,greenbaum2012maskless,sobieranski2015portable}. This requires systems to provide precise mechanical subpixel displacement as well as very high reliability and robustness. Recently, wavelength scanning is proposed to effectively avoid mechanical subpixel displacement and require significantly fewer measurements without sacrificing performance \cite{luo2016pixel}. Yet, it needs extra wavelength calibration and dispersion compensation simultaneously that will increase the complexity and accumulate error in the actual operation. In addition, the introduction of the wavelength-tunable light source will increase the cost of the whole system. As for super-resolution image reconstruction in lensfree in-line holographic systems, the super-resolution methods are usually associated with phase retrieval methods, such as the objective-support based single intensity measurement \cite{marie1979digital,koren1993iterative}, the Gerchberg-Saxton algorithm based multiple intensity measurements \cite{greenbaum2012maskless}, the synthetic aperture based multiple angles measurements \cite{luo2016pixel,luo2015synthetic}, the transport of intensity equation (TIE) based intensity measurements on different axial planes \cite{spence2001lensless,zuo2015lensless}. Moreover, these super-resolution reconstruction methods perform pixel super-resolution and phase retrieval sequentially, and usually require considerable quantities of data collections.

In the recent work, Luo \emph{et al} proposed a propagation phasor approach, which for the first time combines phase retrieval and pixel super-resolution into a unified mathematical framework \cite{luo2016propagation}. The approach gives the deduction that besides twin image elimination, the diversity of the sample-to-sensor distance can also be used for aliasing signal elimination. Furthermore, this method significantly simplifies the operation process and improve data efficiency. Meanwhile, the enhancement of resolution only by Z-scanning has been demonstrated \cite{luo2016propagation,wang2016computational}. This method depends on the perfect fit between the practical and theoretical imaging model, but it is difficult to achieve in actual operation. Thus, the experimental reconstructions are not as perfect as the simulation results. On the one hand, the frequently used single registration method to correct the inevitable lateral positional error resulting from Z-scanning is far from meeting the need of recent super-resolution-imaging applications due to the tiny residual error or the mis-registration. On the other hand, due to the non-negligible noise, the stability and quality of the reconstructions will be degraded which is often attributed to the non-convex nature of phase retrieval. Moreover, there is a serious risk that the calculated intensity patterns on different planes have no tendency to be consistent with those in the next iteration during the successive iterative reconstruction process.

In this paper, we propose a method called adaptive pixel-super-resolved lensfree imaging (APLI), in which the iterative phase retrieval method is used to bring forward this super-resolution reconstruction method to overcome or at least partially alleviate above-mentioned limitations. Different from the traditional reconstruction method based on multi-height intensity measurements, the presented method has taken pixel binning process into account. Additionally, this method simultaneously achieves phase retrieval and super-resolution reconstruction only based on a stack of out-of-focus images during the iterative process in the spatial domain. During the reconstruction process, to find optimum solution to the phase retrieval problem and effectively reduce impact of unpredictable disturbance during image acquisition, APLI first introduces the adaptive relaxation factor strategy and the automatic lateral positional error correction. This improves the stability and robustness of the reconstruction towards noise as well as retains rapid convergence speed. We demonstrate the success of our approach by reconstructing the USAF resolution target and the stained biological paraffin section of typical dicot root across the FOV of $\sim$29.85 \(m{m^2}\) with only ten intensity images. Based on the ALPI, we achieve half-pitch lateral resolution of $770$ $nm$, surpassing $2.17$ times of the theoretical Nyquist–Shannon sampling resolution limit imposed by the sensor pixel-size ($1.67$ $\mu m$). The full-FOV reconstruction results and few raw holograms make the method to be a very attractive and promising technique for lensfree microscopy under noisy conditions.

\section*{Materials and Methods}
\subsection*{Experimental Setup}

\begin{figure}[htb]
\centering
\includegraphics[width=13cm]{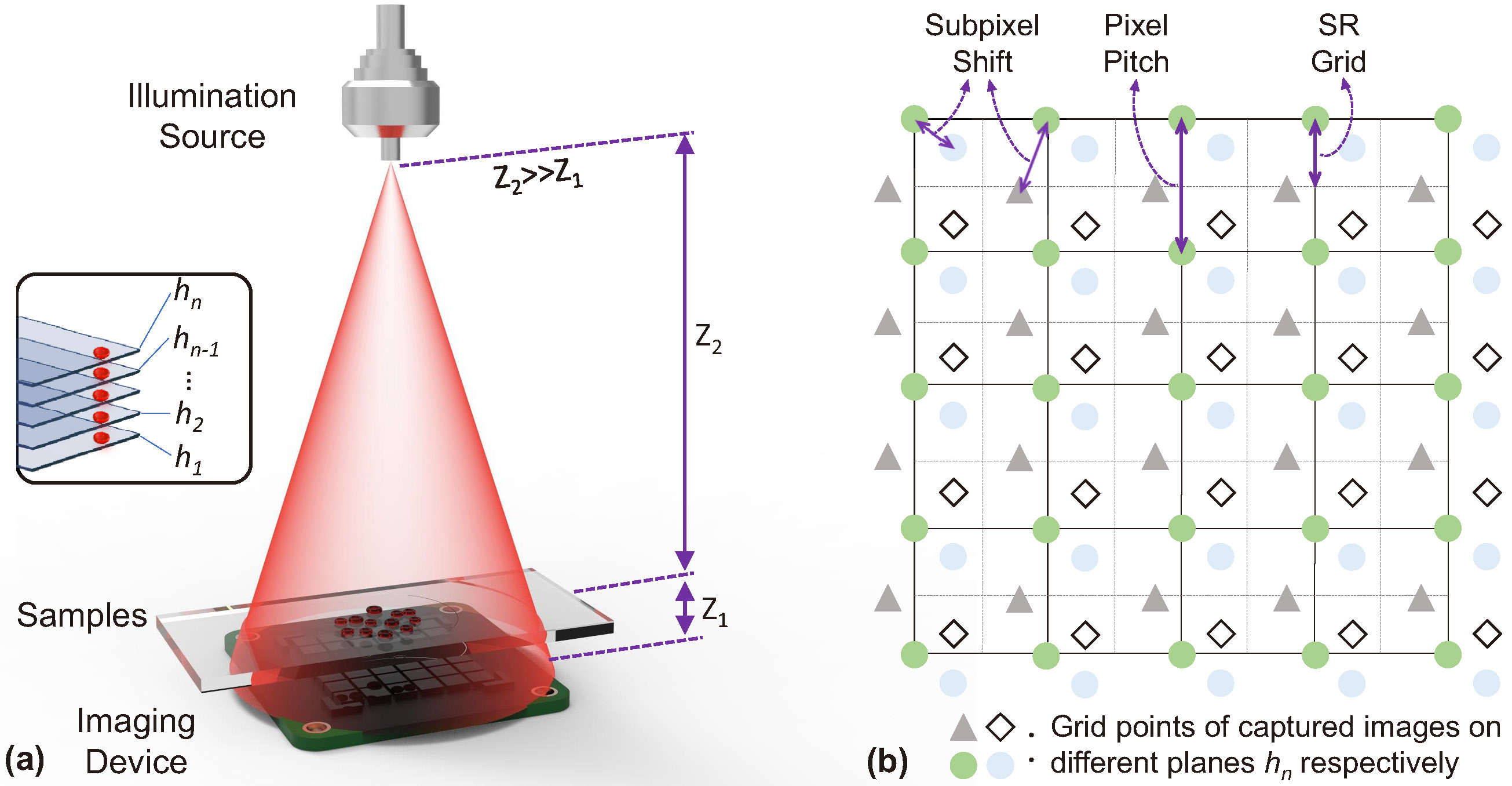}
\caption{(a) Schematic diagram of our setup. The upper left inset shows the multiple intensity measurements on different sample-to-sensor planes. (b) A pictorial example of intensity images at multiple height, in which the subpixel shift results from lateral positional error.}
\label{fig:Fig1}
\end{figure}

Figure \ref{fig:Fig1}(a) depicts the configuration of the lensfree imaging setup. The coherent or partially coherent light irradiates the specimen, and then the scattered light and the transmitted light co-propagate in the same direction, finally forming interference fringes on the imaging device. To achieve wide FOV and reduce the unacceptable resolution loss of the reconstructed images, the sample plane must be much closer to the sensor than the source \cite{oh2010chip} and the length scale of the distance between samples and the imaging device typically is on the order of submillimeter. In order to make the light on the sample plane to be considered as a plane wave when the emitted light impinging on the object plane, the distance between the source and samples should be large. Based on this, the whole active area of the imaging sensor can be regarded as the FOV while the magnification (\(F\)) approaches unit [see Fig. \ref{fig:Fig1}(a), \({Z_2} >  > {Z_1}\) and \(F = ({Z_1}{\rm{ + }}{{\rm{Z}}_2})/{Z_2} \approx 1\)]. Thus, the scale of the FOV will be restricted by the number of pixels of the imaging device, and at the meantime the spatial resolution will be directly influenced by the pixel-size. Unfortunately, during the process of Z-scanning which is implemented to achieve super-resolution, whether electric or manual movement of samples will result in the tiny lateral positional error inevitably and the co-propagated light beam carrying the error will be sampled by the imaging device. As shown in Fig. \ref{fig:Fig1}(b), the tiny error will result in the subpixel shift among the intensity images on the different planes.

As depicted in Fig. \ref{fig:Fig1}(a), our lensfree in-line digital holographic imaging system mainly contains three parts: a single mode fiber-coupled light source (LP660-SF20, Thorlabs, the United States), a monochrome imaging device (DMM 27UJ003-ML, the imaging source, Germany), and thin specimen placed above the imaging device. In our experimental system, the optical fiber is put at $\sim$20 \({cm}\) over the samples. At the meantime, the imaging sensor is placed $\sim$400-900 \(\mu m\) away from the sample which is attached to a positioning stage (MAX301, Thorlabs, the United States). The stage holds the specimen with the self-designed 3D-printed support and is used for the motion of specimen along the direction of Z axes. The camera in our lensfree in-line digital holographic imaging system has 1.67 \(\mu m\) pixel-size and 10.7 megapixels, and it is used for the acquisition of several holograms with ten sample-to-sensor distances \({Z_1}\). Theoretically, the FOV can reach $\sim$29.85 \(m{m^2}\) and simultaneously the resolution is up to 2.5 fold of camera pixel resolution. Finally, the spatial resolution of the experimental results is enhanced to the $2.17$ times of the theoretical Nyquist–Shannon sampling resolution and the space–bandwidth product (Megapixel) is increased from 10.7 to 50.34.

\subsection*{Sample preparation}
A standard \(2'' \times 2''\) positive 1951 USAF resolution test target (Edmund Scientific Corporation, Barrington, New Jersey, USA) is used to quantitatively demonstrate resolution improvement. Besides the resolution target, the typical dicot root (Carolina Biological Supply Company, Burlington, North Carolina, USA) stained with fast green and the counterstain safranin shows general plant anatomy for the study of the internal structure of plants.

\subsection*{Adaptive pixel-super-resolved lensfree imaging (APLI)}
\begin{figure}[htb]
\centering
\includegraphics[width=15cm]{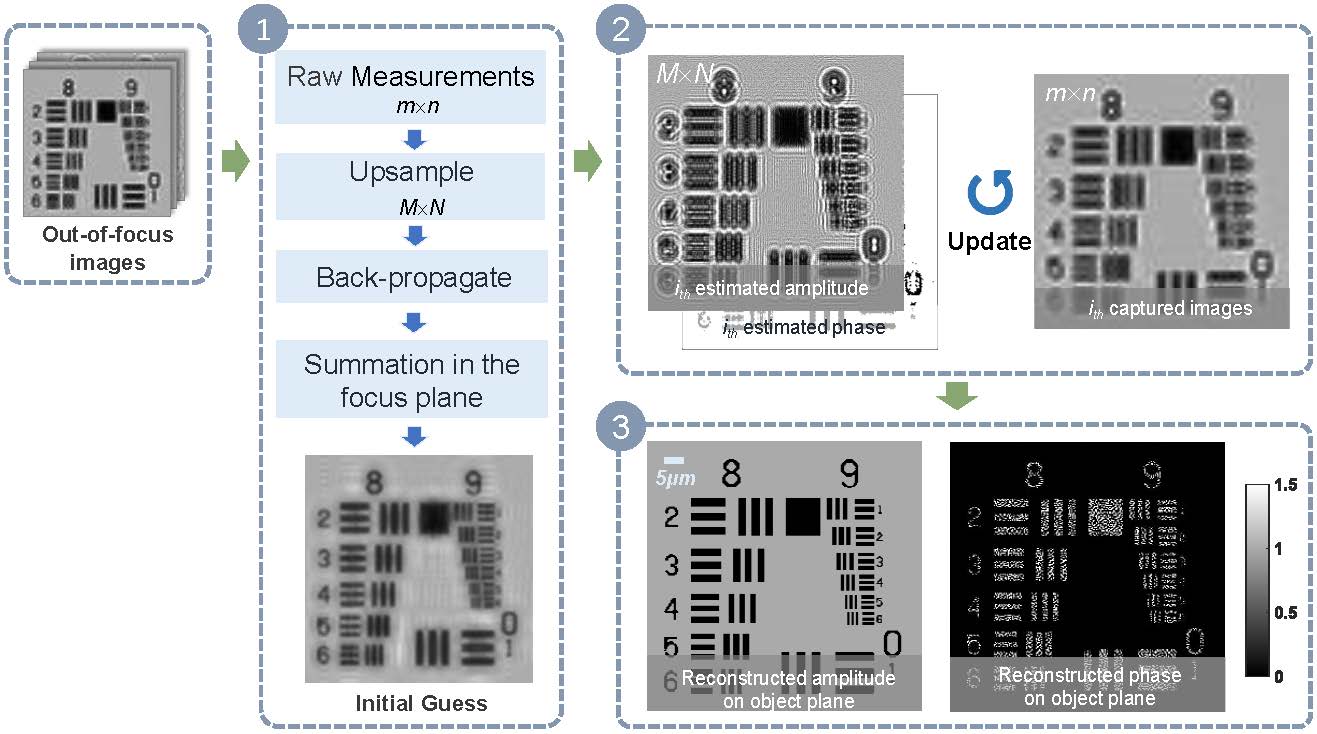}
\caption{Schematic diagram of APLI.}
\label{fig:Fig2}
\end{figure}

In order to recover a super-resolution image of the complex object field based on a series of the out-of-focus low-resolution holograms in the spatial domain, the overview flowchart of our method can be seen in Fig. \ref{fig:Fig2} and described as the following three stages.
\subsubsection*{Stage 1: generation of an initial guess}

A stack of the holograms (\emph{e.g}, the pixel dimension of the hologram is \(m \times n\)) is captured on different sample-to-sensor planes and the first plane should get as close as possible to the sensor. After capturing the raw images, up-sampling will be carried to all holograms with the nearest neighbor interpolation which coincides with the imaging theory of cameras, and all the up-sampling images [\emph{e.g}, the pixel dimension of the hologram is \(M \times N\) with the interpolation weight $k$ ($M \times N = km \times kn$)] will back-propagate to the object plane with the auto-focusing algorithm\cite{mudanyali2010detection,pech2000diatom}. All the up-sampling intensity images are superimposed together to acquire a good initial guess which will be used as the input of \textbf{\emph{Stage 2}}.

Although the single back-propagated up-sampling hologram can be regarded as the initial guess, simply summing up all back-propagated up-sampling holograms can significantly suppress the twin image noise, aliasing signals and up-sampling related artifacts \cite{luo2016propagation,wang2016computational}. Furthermore, with the same set of raw data, this initialization method can have resolution improvement of the initial guess compared to the previous initialization method \cite{haddad1992fourier} and then have faster convergent rate in the \textbf{\emph{Stage 2}}.
%\textbf{\emph{Stage2:}}
\subsubsection*{Stage 2: iterative multi-height images reconstruction}
The whole iteration process is a procedure for phase retrieval, and it is essentially a process of solving the inverse problem which is very common in the computational imaging. To solve this problem, we need to build a precise forward model and reconstruct the super-resolution intensity images and the phase map from the captured discretized intensity images. and then the assumptions of super-resolution intensity distribution and phase map (as mentioned in the above subsection) are input into the precise forward model to obtain the estimated captured images. If the estimated captured images can accord with the actual captured images, the assumptions will be regarded as the actual super-resolution intensity and phase images.

Thus, the following two key elements must be taken seriously to make the model to be consistent with actual physical process. Firstly, in order to obtain the precise imaging model, the limited sensor pixel-size and the unpredictable disturbance during image acquisition should be taken into account. The pixel binning as the process of recording the image is a down-sampling procedure which can be regarded as the spatial averaging. On the other hand, to acquire the diffraction patterns on the distinct planes, the longitudinal shift of the samples is inevitable which will lead to the accidental lateral displacement. Hence, the lateral positional error must be absorbed in the model. (More details are given in the Section \textbf{\emph{Automatic lateral positional error correction}}.) Secondly, the stability and the robustness of the solution to the phase retrieval problem must be improved, and the algorithm should be able to converge to a desired optimal solution that can be considered as the optimization of phase recovery based on multi-height measurements. Solving the optimization problem is deemed to make the current estimate close fit the input actual captured images as a whole, and the quantification is given by the real-space error as described in the following equation \begin{equation}
\varepsilon  = {\sum\nolimits_i {\left\| {\sqrt {{{{I}}_i}}  - |{{{g}}_i}|} \right\|} ^2}
\end{equation}
where ${\left\| . \right\|}$ is the Euclidean norm, $\sqrt {{{{I}}_i}}$ is the amplitude of ${i_{th}}$ captured image, $|{{{g}}_i}|$ is the current down-sampling estimated amplitude corresponding to the ${i_{th}}$ measurement, which is the output after consideration of system uncertainties. The solution process is an incremental gradient optimization process instead of the traditional one, which will provide a relatively correct solution in early iterations, but then overshoot. This problem is often attributed to the non-convex nature of phase retrieval, but we find the reason for this is more closely related to the choice of the relaxation factor based on the analysis of the similar problem in pre-work \cite{zuo2016adaptive}, and the relaxation factor needs to be gradually diminishing for convergence even in the convex case. Thus, the adaptive relaxation factor should be introduced into the model to achieve the improvement in the stability and robustness of the reconstruction towards noise. (Details can be referred to the Section \textbf{\emph{Adaptive relaxation factor}}.)

\begin{figure}[!b]
\centering
\includegraphics[width=14cm]{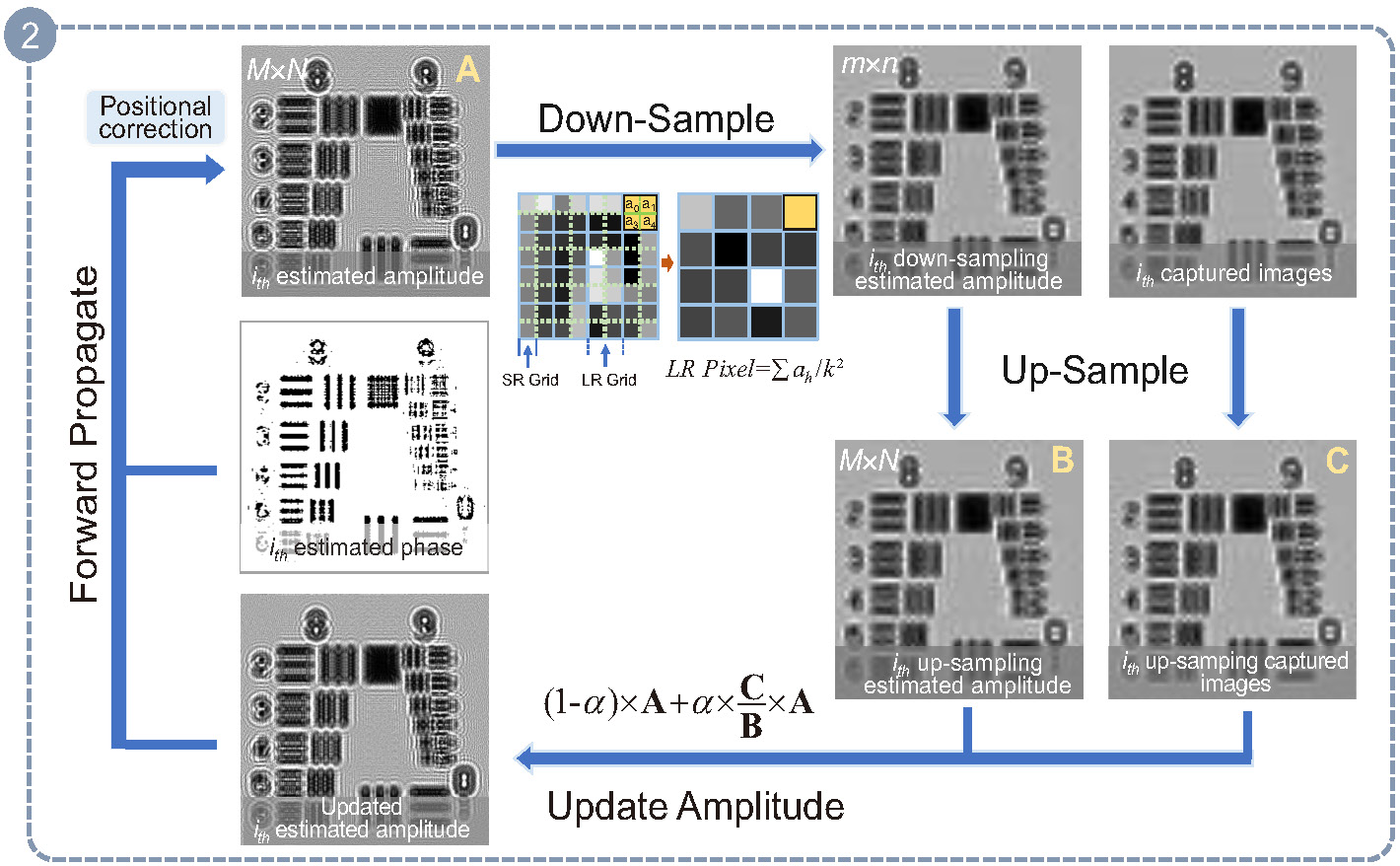}
\caption{The flow diagram of one iteration cycle. \textbf{A}, \textbf{B} and \textbf{C} represent the matrix of the ${i_{th}}$ estimated amplitude, ${i_{th}}$ up-sampling estimated amplitude, and ${i_{th}}$ up-sampling amplitude of the captured image, respectively.}
\label{fig:Fig3}
\end{figure}

The updating amplitude step is crucial and depends on a correction coefficient matrix. This matrix is determined by the product of the adaptive relaxation factor \(\alpha\) and the proportional relation matrix between the up-sampling captured images and the prior estimated intensity images. The specific process can be seen in Fig. \ref{fig:Fig3} and described as the following three steps.

\textbf{\emph{Step 1}}, the (\emph{i-1})\({_{th}}\) estimated complex amplitude {\(O_{i - 1}^j\)} is forth-propagated to the next height ({\(O_{i}^j\)}) and then the \({i_{th}}\) captured images are up-sampled (\emph{j} represents the current index of the iteration cycle and \emph{i} represents the index of the out-of-focus plane.). Next, the estimated intensity image \(|O_i^j{|^2}\) is registered with the up-sampling captured images ${I_{upsample}}$, and the positional error is represented by $({x_{shift}},{y_{shift}})$. Then we shift \(|O_i^j{|^2}\) in place by the amount of $({-x_{shift}},{-y_{shift}})$ and the original estimated intensity image is substituted by the refined intensity termed $|O_{i\_ref}^j{|^2}$ (the amplitude can be denoted as \textbf{A}).

\textbf{\emph{Step 2}}, we implement down-sampling to the refined estimated super-resolution intensity images $|O_{i\_ref}^j{|^2}$ with the point spread function (PSF) of the low-resolution sensor as shown in the left upper Fig. \ref{fig:Fig3}. PSF is usually modeled as a spatial averaging operator $LR{\kern 1pt} {\kern 1pt} Pixel = \frac{{\sum {{a_h}} }}{{{k^2}}}(h = 0,1 \ldots ,{k^2} - 1)$ \cite{park2003super}, where ${a_h}$ is the gray value of the super-resolution intensity images, \emph{k} is a down-sampling factor.

\textbf{\emph{Step 3}}, after down-sampling, the estimated low-resolution intensity image has the same dimension with the original captured image on the corresponding \({i_t}\)\({_h}\) sensor-to-sample plane. Then we up-sample the estimated low-resolution intensity image and the corresponding captured image (the amplitude can be referred as the matrix \textbf{B} and \textbf{C} respectively.). To acquire the correction coefficient matrix, we multiply the adaptive relaxation factor \(\alpha\) with proportional value between the matrix \textbf{C} and \textbf{B}, and the expression $(1 - \alpha ) \times {\mathbf{A}} + \alpha  \times \frac{{\mathbf{C}}}{{\mathbf{B}}} \times {\mathbf{A}}$ will be regarded as the updated \({i_{th}}\) estimated amplitude. The relaxation factor \(\alpha\) in above expression is a diminishing value differing from the traditional fixed value $\sim$0.5, and the guided filter is taken into account to further eliminate the influence of noise as well. At last, the complex amplitude containing the new updated estimated amplitude and the previous unchanged phase is forth-propagated to the next height using the Angular Spectrum Method \cite{goodman2005introduction}.

The process of the \textbf{\emph{Step 1}}-\textbf{\emph{Step 3}} is repeated until all the sample-to-sensor distances are gone through. That is to say, all the raw measurements are used for once, and it will be considered as one iteration cycle.

\subsubsection*{Stage 3: reconstruction on the object plane}
After some iterations, we will achieve the complex amplitude on the plane closest to the imaging device and then back-propagate the complex amplitude to the object plane as shown in Fig. \ref{fig:Fig2}.

\subsection*{Physical modeling of the pixel binning}

\begin{figure}[!htb]
\centering
\includegraphics[width=11cm]{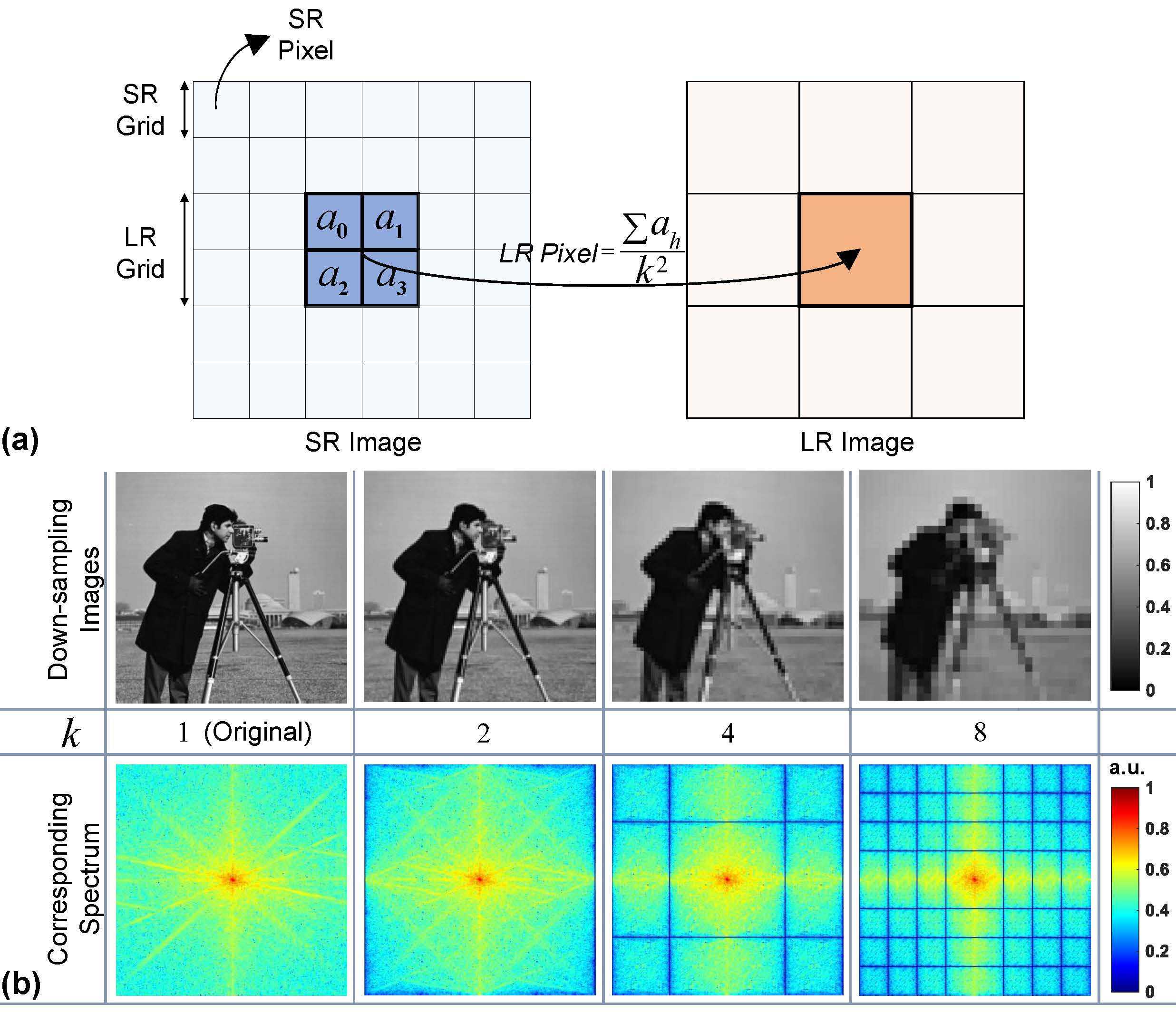}
\caption{(a) The PSF of the low-resolution sensor. (b) The natural loss of spatial resolution after sampling.}
\label{fig:Fig4}
\end{figure}

Blurring may be caused by an optical system (inherent noise of the camera, diffraction limit, \emph{etc}.), and the PSF of the imaging device. The former can be modeled as linear space invariant while the the latter is considered as linear space variant \cite{park2003super}. It is difficult to obtain the exact information about the linear space invariant, so it is usually compensated by the specific algorithms or avoided as much as possible. Besides the linear space invariant, in the process of image reconstruction, the PSF of the imaging device (which can also be regarded as the finiteness of the physical pixel-size) is an important factor for blur, which should be incorporated into the reconstruction procedure. As a complementary interpretation, there is a natural loss of spatial resolution caused by the insufficient sensor density and noise that occurs within the sensor or during transmission. As shown in Fig. \ref{fig:Fig4}(b), the spectrum loss will be more serious while the decimation factor increases. Figure. \ref{fig:Fig4}(b) indicates again that the pixel-size is the main limiting factor of the systems which will determine whether it can directly record the high frequency fringes corresponding the super-resolution of the samples.

In traditional multi-height reconstruction method \cite{greenbaum2012maskless,gerchberg1972practical}, many efforts are made to implement subpixel shift to achieve the super-resolution, but the pixel binning is not taken into account, which does not accord with actual physical process. Thus, involving the process of recording digital images in the reconstruction procedure has drawn attention, and the enhancement in resolution has been validated \cite{luo2016propagation}. The process of recording digital images is a down-sampling process which is usually modeled as a spatial averaging operator [\emph{LR} \(Pixel = \frac{{\sum {{a_h}} }}{{{k^2}}}(h = 0,1...{k^2} - 1)\) where \({a_h}\) is the gray value of the super-resolution intensity images, \emph{k} is a decimation factor] as shown in Fig .\ref{fig:Fig4}(a). In the iterative process of the reconstruction, we convolve the estimated intensity of the field $|O_i^j{|^2}$ with the PSF of the image sensor \cite{park2003super}, and it has the same dimension as the raw measurement.

\subsection*{Automatic lateral positional error correction}
In many lensfree systems, the different sample-to-sensor distances are provided by the mechanical movement, so a lot of mechanical error will be generated as shown in Fig. \ref{fig:Fig1}(b). In order to eliminate the unavoidable error at subpixel-scale, the traditional method is registering images before reconstruction \cite{greenbaum2012maskless,sobieranski2015portable} called beforehand lateral positional error correction (BLPEC). However, in the actual imaging process, the light illuminates the specimen which has tiny lateral movement because the positioning stage will lead into the lateral mechanical error while it moves longitudinally, and then forth-propagates to the imaging plane carrying the information of the object. Thus, the lateral positional error appears before down-sampling and then causes the lateral positional error of the captured images. Based on this, the BLPEC can only correct the lateral positional error cursorily, and the accuracy of the correction will decrease due to existence of the artifacts and aliasing. Additionally, this method has incapacity to rectify registration error in the later process which will affect the final quality of reconstruction.

In order to solve the problems existed in traditional BLPEC, we introduce automatic lateral positional error correction (ALPEC) into our method. The crux of solving the general problem of subpixel image registration is computing the cross correlation between the image to register and a reference image by means of a fast Fourier transform (FFT), and locating its peak \cite{guizar2008efficient}. The cross correlation of the captured image $f(x,y)$ and its corresponding estimated image $g(x,y)$ is defined by:
\begin{equation}
\begin{gathered}
  {r_{fg}}({x_{shift}},{y_{shift}}) = \sum\limits_{x,y} {f(x,y){g^*}(x - {x_{shift}},y - {y_{shift}}} ) \hfill \\
  {\kern 42pt} = \sum\limits_{u,v} {F(u,v){G^{\text{*}}}(u,v} )\exp \left[ {i2\pi (\frac{{u{x_{shift}}}}{M} + \frac{{v{y_{shift}}}}{N})} \right] \hfill \\
\end{gathered}
\end{equation}
where \emph{N} and \emph{M} are the image dimensions, (*) represents complex conjugation, \emph{F} and \emph{G} denote the discrete Fourier transform of the \emph{f} and \emph{g} respectively. The expression of $F(u,v)$ is $F(u,v) = \sum\limits_{x,y} {\frac{{f(x,y)}}{{\sqrt {MN} }}\exp \left[ { - i2\pi \left( {\frac{{ux}}{M} + \frac{{vy}}{N}} \right)} \right]}$ and there is a similar expression for $G(u,v)$. It is important to determine accurately the peak of the cross-correlation function ${r_{fg}}({x_{shift}},{y_{shift}})$, and relax the limitation on computational speed and memory caused by the FFT. Thus, the refined initial estimate method \cite{guizar2008efficient,fienup1997invariant} with aid by the existence of analytic expressions for the derivatives of ${r_{fg}}({x_{shift}},{y_{shift}})$ with respect to ${x_{shift}}$ and ${y_{shift}}$ is used, and the algorithm iteratively searches for the image displacement $({x_{shift}},{y_{shift}})$ that maximizes ${r_{fg}}({x_{shift}},{y_{shift}})$. At last, it can achieve registration precision to within an arbitrary fraction of a pixel at a fast rate.

\subsection*{Adaptive relaxation factor}
In most cases, the propagation phasor approach \cite{luo2016propagation} is an effective solution to the pixellation problem and it gives a unified mathematical framework combining phase retrieval and pixel super-resolution. Nevertheless, in practical operation, the stability and reconstruction quality of the method may be significantly degraded due to the existence of nonnegligible noise paralleling down-sampling in the captured images. This problem is often attributed to the non-convex nature of phase retrieval and the ill-condition process of the super-resolution reconstruction. Although numerous super-resolution algorithms have been proposed in the literature \cite{bishara2010lensfree,greenbaum2012maskless,luo2016pixel,huang1984multi,baker2002limits,yang2008image}, the super-resolution image reconstruction remains extremely ill-posed \cite{baker2002limits,yang2008image}. What’s worse, the noise effect will accumulate as the iterations augment.

The choice of the relaxation factor can suppress noise to a certain extent, typically the relaxation factor \(\alpha  = 0.5\) in traditional phase retrieval methods. The relaxation factor will be utilized to update amplitude, and the corrected super-resolution intensity images substitutes for the earlier estimated super-resolution intensity images incompletely. In other words, the corrected super-resolution intensity images will occupy a part in the new updated estimated intensity (as shown in Fig. \ref{fig:Fig3}). The corrected super-resolution intensity images have a close relationship with the captured intensity images containing noise, so the new updated estimated intensity images suffer from the noise because the captured intensity images will impose ill-condition restrictions with the fixed relaxation factor.

The stability and reconstruction quality may be significantly degraded when non-negligible noise is present in the captured images, and the same problem may be encountered in FPM. We find that the reason for the phenomenon in this field is the non-convex nature of phase retrieval and more closely related to the choice of the step-size, so the adaptive step-size strategy is introduced to successfully solve this problem \cite{zuo2016adaptive}. Considering that the problems of the iterative method in lensfree imaging are also attributed to the non-convex nature of phase retrieval, instead of the traditional fixed relaxation factor, the adaptive relaxation factor which diminishes to an infinitely small value will be used to improve the performance of the the incremental solutions. So the critical issue in practical application will be how to determine a suitable relaxation factor sequence ${\alpha ^{iter}}$ to get close to a solution within fewer iterations. The ${\alpha ^{iter}}$ must satisfy the two conditions that are shrinking the relaxation factor to zero and making the diminishing speed not be too fast. Because if the relaxation factor shrinks too fast, the estimated object field may converge to a point that isn't a minimum especially when the initial point is sufficiently far from the optimum. So the relaxation factor should not be reduce too fast and then the algorithm can travel infinitely far. Thus, in this paper, we give an alteration of the relaxation factor when the global error \(\varepsilon ({{\mathbf{O}}^{iter - 1}})\) and $\varepsilon ({{\mathbf{O}}^{iter}})$ obtained in consecutive cycles satisfies the following criterion:
\begin{equation}
{\alpha ^{iter}} = \left\{ \begin{gathered}
  {\alpha ^{iter - 1}}{\kern 80pt} otherwise \hfill \\
  {\alpha ^{iter - 1}}/2{\kern 30pt} [\varepsilon ({{\mathbf{O}}^{iter - 1}}) - \varepsilon ({{\mathbf{O}}^{iter}})]/\varepsilon ({{\mathbf{O}}^{iter}}) < \eta  \hfill \\
\end{gathered}  \right.
\end{equation}
where ‘\emph{iter}’ is the index of the iteration cycle, ‘\(\eta \)’ is a small constant which should be much less than 1. The global error $\varepsilon ({{\mathbf{O}}^{iter}})$ is determined by $\varepsilon ({{\mathbf{O}}^{iter}}) = {\sum\nolimits_i {\left\| {\sqrt {{{\mathbf{I}}_i}}  - |{{\mathbf{g}}_i}|} \right\|} ^2}$, where $\left\| . \right\|$ is the Euclidean norm. The captured image and the estimate function is raster-scanned into vectors ${{\mathbf{I}}_i} = {\{ {I_i}\} ^{m \times n}}$  and ${{\mathbf{g}}_i} = {\{ {g_i}\} ^{m \times n}}$ (with $m \times n$ pixels). ${g_i} = {\mathbf{O}}_i^{iter} \otimes {\text{PSF}}$ is spatial averaging operation, and ${\mathbf{O}}_i^{iter}$ is an ${M \times N}$ matrix (the estimated intensity of the field) while PSF is determined by the intrinsic property of the camera. Here we should know that the $|{{\mathbf{g}}_i}|$ consider the system uncertainties such as the lateral positional error. Finally, the algorithm will converge to the stationary point when the relaxation factor reach a pre-specified minimum.

%\newpage

\section*{Discussion and Results}

\subsection*{The comparison between the adaptive and fixed factor}

\begin{figure}[!htb]
\centering
\includegraphics[width=14.2cm]{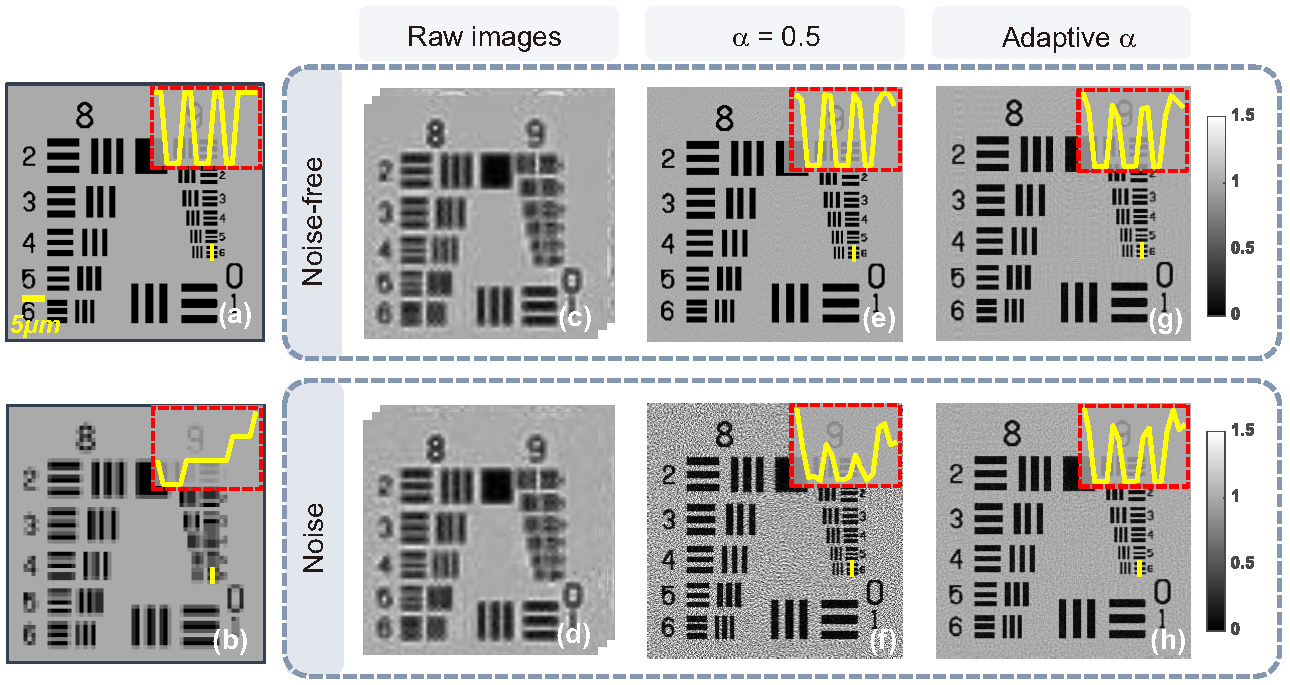}
\caption{ The impact of the noise on the system and the noise-restraining with an adaptive relaxation factor in iterative process. (a) The super-resolution image needed to recover in theory, (b) the low-resolution image captured by the camera on the object plane theoretically, (c)-(d) the images captured by the camera on the planes of distinct sample-to-sensor distances with noise and not respectively, (e)-(f) the reconstructed super-resolution images using a fixed relaxation factor ($\alpha  = 0.5$) under the noise-free and noisy circumstances separately, (g)-(h) the reconstructed super-resolution images using the adaptive relaxation factor under the noise-free and noisy circumstances separately.}
\label{fig:Fig5}
\end{figure}

Figure \ref{fig:Fig5} shows the influence of the noise on the system and emphasizes the important role played by the relaxation factor in iterative process. A theoretical super-resolution image needed to reconstruct is shown in Fig. \ref{fig:Fig5}(a), and Fig. \ref{fig:Fig5}(b) shows the low-resolution image captured by the camera on object plane in theory. Figures. \ref{fig:Fig5}(d) and \ref{fig:Fig5}(c) describe a set of images captured by the camera on the planes of distinct sample-to-sensor distances in our simulation with Gaussian noise and not respectively. Figures. \ref{fig:Fig5}(e) and \ref{fig:Fig5}(f) depict the reconstructed super-resolution images using the fixed relaxation factor $\sim$0.5 under the noise-free and noisy circumstances separately. The two yellow curves in the sub-graphs of red region convey the information that if we update the earlier estimated super-resolution intensity images using a fixed relaxation factor (\(\alpha  = 0.5\)) as the new estimated amplitude in the above-mentioned Section \textbf{\emph{APLI}}, the reconstructed result with noise [see Fig. \ref{fig:Fig5}(f)] compared to the noise-free reconstructed output [see Fig. \ref{fig:Fig5}(e)] is worse in respect of the resolution and background after the same iterations. To demonstrate that an adaptive relaxation factor can effectively solve the problem having a close relationship with the over-amplification noise, we test our method under the noise-free and noisy circumstances separately. The results can tell apart the densest line and give relatively clean background under the two different conditions are shown in Figs. \ref{fig:Fig5}(g) and \ref{fig:Fig5}(h) respectively.

\begin{figure}[!htb]
\centering
\includegraphics[width=15cm]{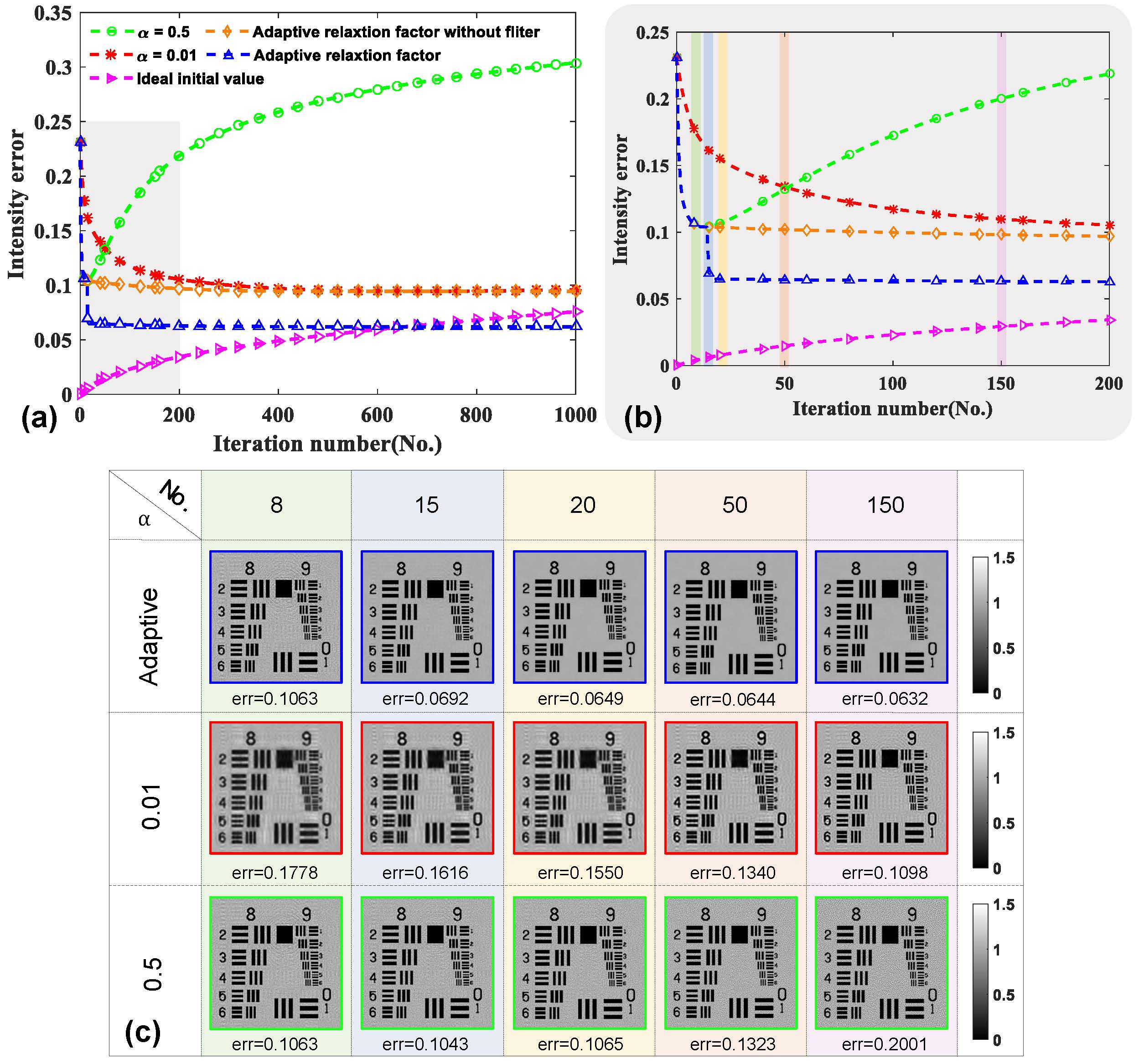}
\caption{The quantitative comparison of reconstruction accuracy versus intensity noise among using the adaptive and fixed relaxation factors, as well as the rate of convergence.}
\label{fig:Fig6}
\end{figure}

Furthermore, the quantitative comparison of reconstruction accuracy versus intensity noise among using the adaptive and fixed relaxation factors (\(\alpha  = 0.5,0.01\)), as well as the rate of convergence is shown in Fig. \ref{fig:Fig6}. Figure \ref{fig:Fig6}(c) depicts the reconstructed results corresponding to iterations labelled in Fig. \ref{fig:Fig6}(b) under the condition of the adaptive and fixed relaxation factors respectively. Figure \ref{fig:Fig6}(a) shows the curves of the intensity error following the iterations increasing with different relaxation factors and Fig. \ref{fig:Fig6}(b) shows the local enlarged drawing of Fig. \ref{fig:Fig6}(a) (shaded region). Among the curves, the purple bight represents that using the perfect initial guess and a very small fixed relaxation factor $\sim$0.01, the intensity error still accumulates as the iterations increase. This offers an explanation for the following phenomenon that with the fixed relaxation factors, the reconstruction error will have convergent tendency at the outset and then get worse after reaching their respective minima which can be obviously seen in the green curve of Fig. \ref{fig:Fig6}(a). The same goes for the small relaxation factor corresponding to the red curve but it is not obvious to observe the the turning point of the curve (the red curve reaches the minimum in the about 700 iterations and then overshoots), because the speed of convergence is extremely slow and the curve rises at a glacial pace after reaching the minimum. The iteration should be suspended when the curves reach their respective minima in succession due to the overshooting of the curve in the later period. The cause of the overshooting is that even if the reconstruction converges to a true value, the captured images still provide the ill intensity constraints as before. Non-convergence is disadvantage for the iterative methods, and suspending the iteration when reaching the minimum will result in loss of image details or taking a long time. Comparing the orange curve with the green one or the red one, we can find that using adaptive relaxation factor can obtain the converged reconstruction and effectively prevent the overshooting. Meanwhile the introduction of the adaptive relaxation factor into our method can retain the relatively fast initial convergence speed and it is seen that this method decreases more rapidly than fixed relaxation factor methods (\(\alpha  = 0.01\)) and converges in the early 20 iterations.

Although the adaptive relaxation factor has the anti-noise capability to a certain extent, the noise will cause the estimated intensity image to have no tendency to be consistent one with another on the next plane and the reconstructed image to deviate from the theoretically calculated values. To avoid the over-amplification of noise we take into account nonlinear denoising algorithm termed guided filter \cite{he2013guided}. It is essentially equivalent to adding the object related prior information that objects are piecewise smooth. The introduction of the guided filter will further restrain the noise and the smooth regions of the reconstruction results will tend to the ideal value, but the reconstruction results are slightly flawed at edge because the guide filter cannot distinguish whether it is noise or jump edges of the object and preserves the edges during the reconstruction process. The combination of the adaptive relaxation factor and the guided filter can effectively suppress the noise and achieve better reconstruction results corresponding to the blue curve. From these results, we can safely conclude that the adaptive relaxation factor method outperforms the fixed relaxation factor methods, with both faster convergence rate and lower mis-adjustment error simultaneously achieved.

\subsection*{The comparison between BLPEC and ALPEC}

\begin{figure}[!htb]
\centering
\includegraphics[width=14cm]{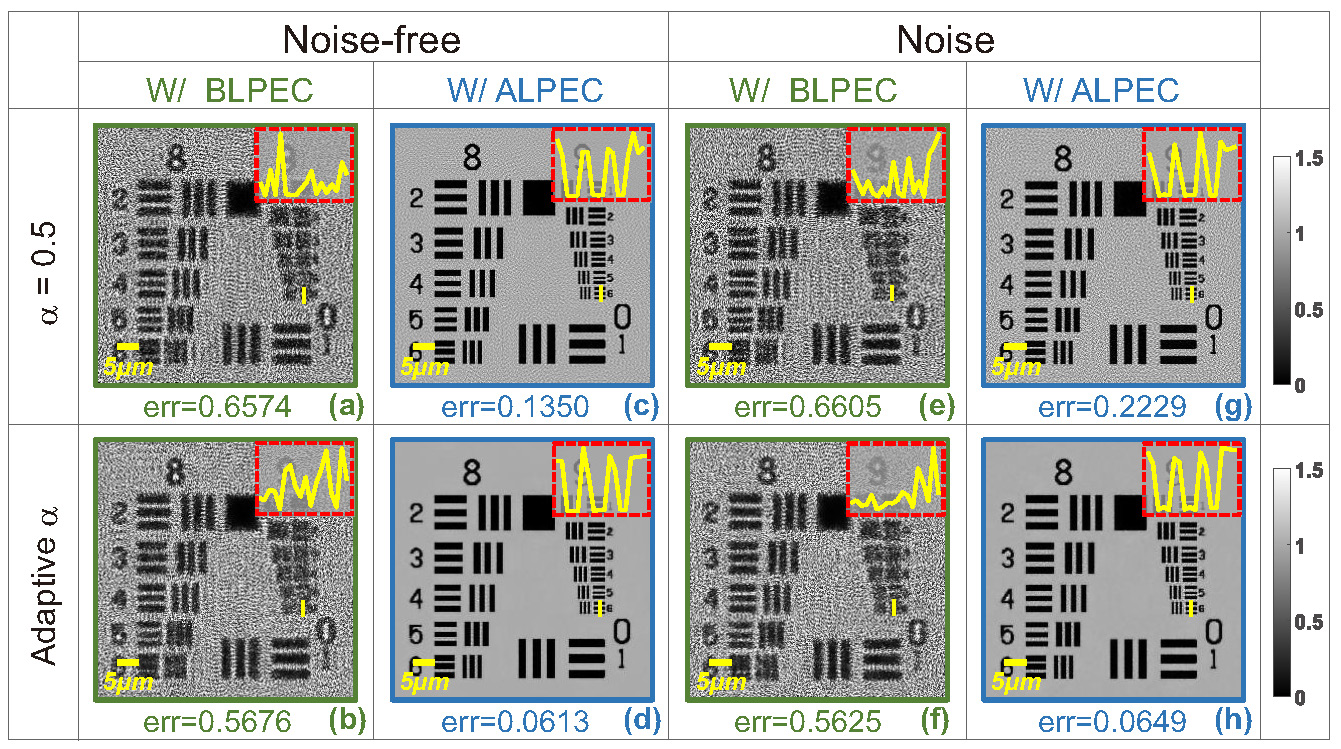}
\caption{The effect of mechanical lateral positional error on the system and great improvement in reconstructed results based on lateral positional error correction.}
\label{fig:Fig7}
\end{figure}

Figure \ref{fig:Fig7} shows that the mechanical lateral positional error has great effect on the systems and lateral positional error correction significantly improves reconstruction performance. The theoretical super-resolution image needed to reconstruct is shown in Fig. \ref{fig:Fig5}(a). Figures. \ref{fig:Fig7}(a)-\ref{fig:Fig7}(h) all have lateral positional error. Figures. \ref{fig:Fig7}(a)-\ref{fig:Fig7}(d) and Figs. \ref{fig:Fig7}(e)-\ref{fig:Fig7}(h) are in clean and noisy environments respectively. To further describe that ALPEC can be widely used in the cases of either the fixed or adaptive relaxation factor, the simulation analysis is conducted as shown in Fig. \ref{fig:Fig7}. Comparing Figs. \ref{fig:Fig7}(a) and \ref{fig:Fig7}(b) with Figs. \ref{fig:Fig7}(c) and \ref{fig:Fig7}(d), we can deduce that the residual tiny lateral positional error can cause serious deviation to the reconstructed results and ALPEC is extremely effective to correct positional error under the condition of either fixed or adaptive relaxation factor but without noise. To achieve more intuitive comparison between whether introducing ALPEC or not, the reconstructed results with noisy captured raw images are shown in Figs. \ref{fig:Fig7}(e) and \ref{fig:Fig7}(f) as well as Figs. \ref{fig:Fig7}(g) and \ref{fig:Fig7}(h) respectively. From above-mentioned comparative simulation, we can find that the remanent lateral positional error brings disastrous distortions to the reconstructed results. Using APLI (with ALPEC), the effect of the lateral positional error and noise will be effectively removed without any prior knowledge.

%\newpage
As shown in the preceding graphs (Figs. \ref{fig:Fig5}-\ref{fig:Fig7}), simulation studies are conducted under the following conditions:

 1) The down-sampling by a factor of four is implemented which can also be regarded as the spatial averaging operating weight of the sensor (the average value of sixteen pixels in the super-resolution image is equivalent to the value of the corresponding pixel in the low-resolution image). Here we should know that the actual physical phenomena and process is that the super-resolution image propagates in the free space, and then the two-dimensional continuous intensity distribution of the hologram (diffraction patterns) is discretized into a matrix through the two-dimensional convolution of the hologram and a pixel unit in an imaging array, which results in the low-resolution image.

 2) In each group of simulations, variable-controlling approach (for instance, the number of iteration remains unchanged) is used to make the conclusion more convincing.

 3) In every simulation, at least 32 raw low-resolution images (the number of pixels is \(m \times n\)) are utilized to reconstruct super-resolution images [the number of pixels is $M \times N$ ($M = k \times m$, $N = k \times n$, $k = 4$)]. These captured diffraction patterns are enforced as object constraints, gradually converging to the missing two-dimensional phase information \cite{fienup1986phase} and the corresponding super-resolution amplitude. For a complex intensity object function, to obtain the super-resolution intensity, the recovery problem becomes undetermined by a factor of 2 since there are \(2 \times M \times N\) pixels defining the object function (\(M \times N\) pixels for the real part and \(M \times N\) pixels for the imaginary part), whereas there are only \(m \times n\) pixels in the measurement matrix \cite{miao1998phase,miao2000oversampling}. In order to solve this underdetermined problem, more information about the object function needs to be acquired and incorporated as a constraint on the solution space, so at least \(2 \times k \times k\) raw low-resolution images are needed in theory \cite{miao1998phase}.

\subsection*{The experimental results of the USAF resolution test target}

\begin{figure}[!htb]
\centering
\includegraphics[width=14cm]{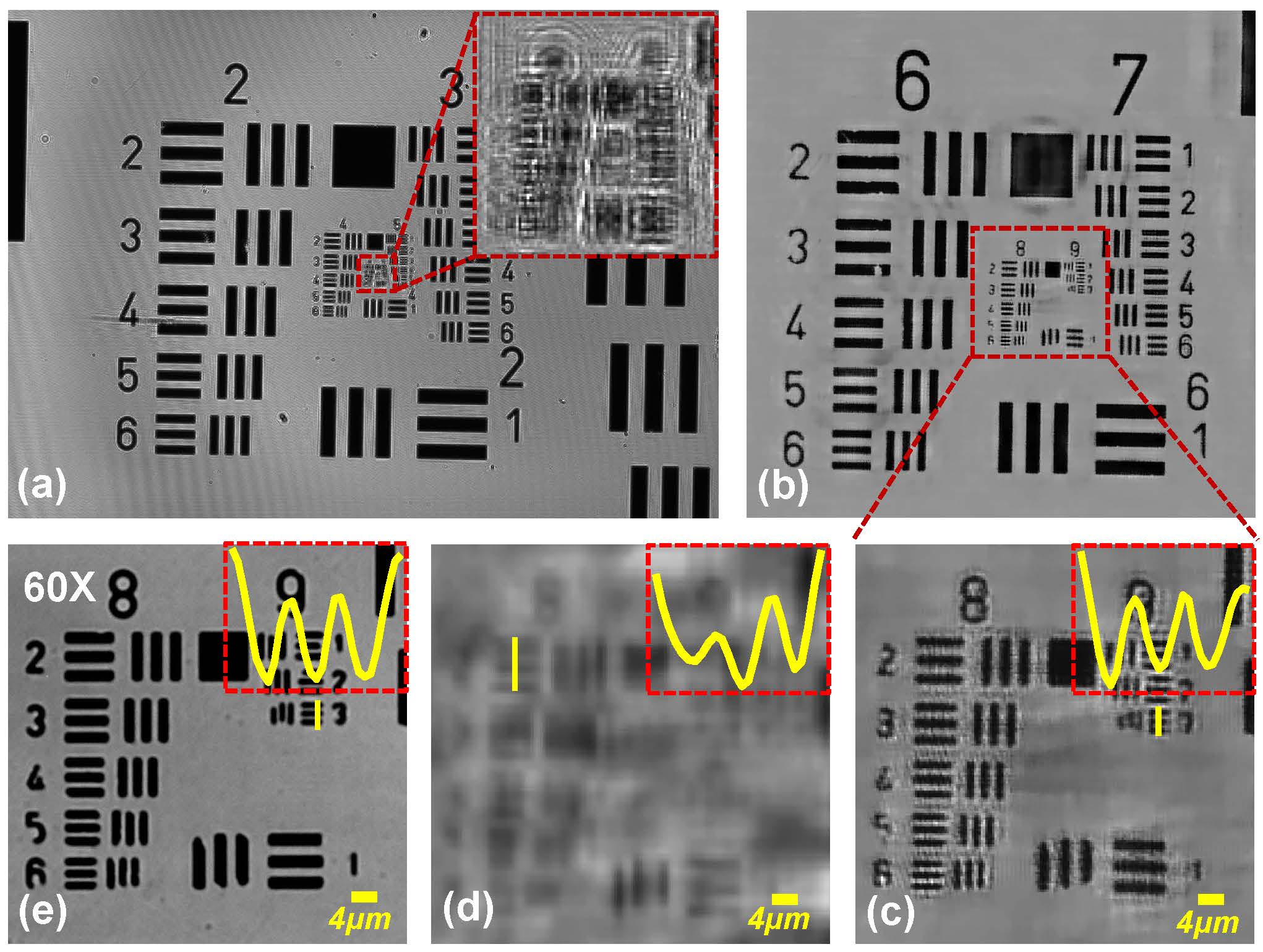}
\caption{The experimental results of the standard 1951 USAF resolution test target. (a) The full FOV low-resolution hologram. (b) The reconstructed intensity image based on our method. The FOV of (b) corresponds to the enlargement of the dashed box of (a). (c) shows the enlarged region corresponding to the red boxed area of (b). (d) The reconstructed intensity image based on the conventional multi-height reconstruction method \cite{greenbaum2012maskless,gerchberg1972practical} (e) The image acquired via plan semiapochromat objective (Olympus, UPLFLN 60XOIPH, NA 1.25-0.65) in 8-bit grayscale range is provided for comparison.}
\label{fig:Fig8}
\end{figure}

A standard 1951 USAF resolution test target as the experimental samples is utilized to prove that our method has the universality and stability during the actual measurements. In order to test our method, we acquire 10 raw holograms at different sample-to-sensor distances ($\sim$547-577 \(\mu m\)) with the standard 1951 USAF resolution test target and each raw hologram is digitized by the imaging device with $1.67$ $\mu m$ pixel-size. Figure \ref{fig:Fig8}(a) shows a full FOV ($\sim$$29.85$ $m{m^2}$) low-resolution hologram which is captured by the camera directly, and the inset shows local enlarged drawing of the dashed rectangular area in Fig. \ref{fig:Fig8}(a). Due to the relatively large pixel-size resulting in down-sampling, our method is applied to diminish the effective pixel-size namely improving the resolution. During the process of reconstruction (the image-processing steps are depicted in the Section \textbf{\emph{APLI}}), the raw holograms are used as the intensity constraints and the recovered super-resolution intensity image is shown in Figs. \ref{fig:Fig8}(b) and \ref{fig:Fig8}(c). In Fig. \ref{fig:Fig8}(c), we can deduce that the smallest resolved half-pitch can reach $0.77$ $\mu m$, which exceeds double the resolution of the result [see Fig. \ref{fig:Fig8}(d)] based on the conventional multi-height reconstruction method \cite{greenbaum2012maskless,gerchberg1972practical} using the same raw images. For comparison, the sample digitalized with the 60X objective is shown in Fig. \ref{fig:Fig8}(e). It is important to emphasize that results shown in Figs. \ref{fig:Fig8}(b) and \ref{fig:Fig8}(c) only base on ten raw images which are obtained without wavelength scanning, subpixel lateral displacement and illumination angles scanning, in other word, we only move the sample along the Z-axis.

In Supplementary Video \textcolor[rgb]{0,0,1}{1}, we show a zooming video of the full-FOV reconstructed images of a USAF target with our method and the traditional method \cite{greenbaum2012maskless,gerchberg1972practical} (with BLPEC) respectively. To process the data in parallel, the large format raw image ($3872 \times 2764$ raw pixels) is divided into 35 portions ($700 \times 700$ raw pixels) for computation. Here, the blocks at the end of each row or each column do not have the same pixels as others and the the adjacent portions have 200 pixels overlap with each other. For the reconstructed image ($2800 \times 2800$ pixels), we cut away 200 pixels at the edge and the adjacent portions introduce a certain degree of redundancy (400 pixels) into our stitching. Thus, no observable boundary is present in the stitched region and the blending comes at a small computational cost of redundantly processing the overlapping regions twice.

\begin{figure}[!b]
\centering
\includegraphics[width=16cm]{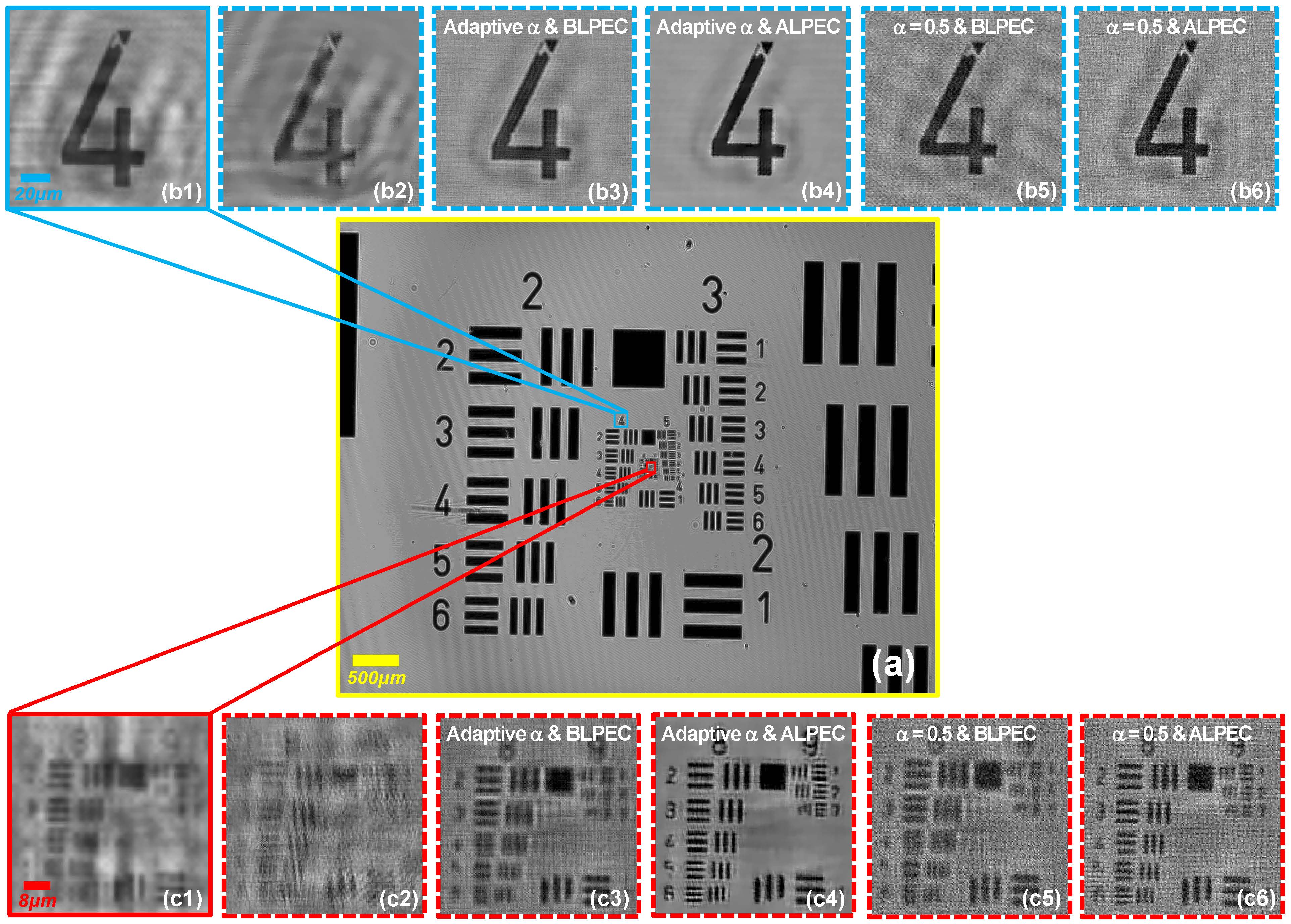}
\caption{The great effects of ALPEC on experimental results, as well as the comparison between the adaptive relaxation and the fixed factor based reconstructed results. }
\label{fig:Fig9}
\end{figure}

Figure \ref{fig:Fig9} describes additional experimental work to address the significance of ALPEC, and intuitively shows the comparison between reconstructed results based on the adaptive and the fixed relaxation factor. Figure \ref{fig:Fig9}(a) presents the FOV of the USAF resolution target recorded by the camera directly.
To illustrate great effects of ALPEC on experimental results, we have carried out two groups of experiments with the adaptive and the fixed relaxation factor respectively. Figures. \ref{fig:Fig9}(b1) and \ref{fig:Fig9}(c1) show the reconstructed results of enlargements of two different small segments in Fig. \ref{fig:Fig9}(a) based on the single raw image, and as shown in them, the reconstructed results are blurry. Figures. \ref{fig:Fig9}(b2) and \ref{fig:Fig9}(c2) show the reconstructed results without positional error correction. From Fig. \ref{fig:Fig9}(b2) we can deduce that the tiny lateral positional error has little effect on low frequency of the reconstructed results, but exerts a tremendous influence on high frequency which corresponds to the super-resolution [see Fig. \ref{fig:Fig9}(c2)]. The same data set is employed to recover the super-resolution image for each segment using our adaptive method with BLPEC and ALPEC respectively. In addition, the same iterations are conducted in two reconstruction methods, the only difference between the methods utilized in this paper is that the latter involves ALPEC while the former puts into effect positional error correction in advance. Figures. \ref{fig:Fig9}(b3) and \ref{fig:Fig9}(c3) present the recovered super-resolution intensity images with BLPEC corresponding to the same segment of Figs. \ref{fig:Fig9}(b1) and \ref{fig:Fig9}(c1) respectively. It can be seen that the silhouette of lines in Fig. \ref{fig:Fig9}(b3) is unsharp but recognizable because the corrected images using BLPEC have removed the relatively large lateral positional error, but still remain tiny the positional misalignment. Even worse, due to the remnant lateral positional error, the higher frequency of the object is still unable to recover shown in Fig. \ref{fig:Fig9}(c3) although the resolution of the reconstructed results has been improved compared to Fig. \ref{fig:Fig9}(c2). With the help of ALPEC, high-quality recovered intensity distributions are obtained, shown in Figs. \ref{fig:Fig9}(b4) and \ref{fig:Fig9}(c4). The blur in Figs. \ref{fig:Fig9}(b3) and \ref{fig:Fig9}(c3) is eliminated completely and the clarity of image is increased, meanwhile better outlines are given in Figs. \ref{fig:Fig9}(b4) and \ref{fig:Fig9}(c4). Similarly, with the fixed relaxation factor (\(\alpha  = 0.5\)), the reconstructed results are shown in Figs. \ref{fig:Fig9}(b5) and \ref{fig:Fig9}(c5) as well as Figs. \ref{fig:Fig9}(b6) and \ref{fig:Fig9}(c6) using the BLPEC and ALPEC separately. We can also come to the same conclusion that the ALPEC can bring benefits to improve the resolution with the adaptive relaxation or fixed factor.

In order to experimentally illustrate the adaptive relaxation factor can improve the stability and robustness of the reconstruction towards noise, the comparison between the adaptive relaxation and the fixed factor based reconstructed results is also shown in Fig. \ref{fig:Fig9}. Figures. \ref{fig:Fig9}(b6) and \ref{fig:Fig9}(c6) show the reconstructed results with fixed relaxation factor (\(\alpha  = 0.5\)) and ALPEC, compared to the reconstructed results using our adaptive relaxation-factor method and ALPEC (the results are shown in Figs. \ref{fig:Fig9}(b4) and \ref{fig:Fig9}(c4)). It is obvious that using the adaptive relaxation-factor method can suppress the over-amplification noise without extra auxiliary information.

\subsection*{Imaging of the typical dicot root}

\begin{figure}[!b]
\centering
\includegraphics[width=14cm]{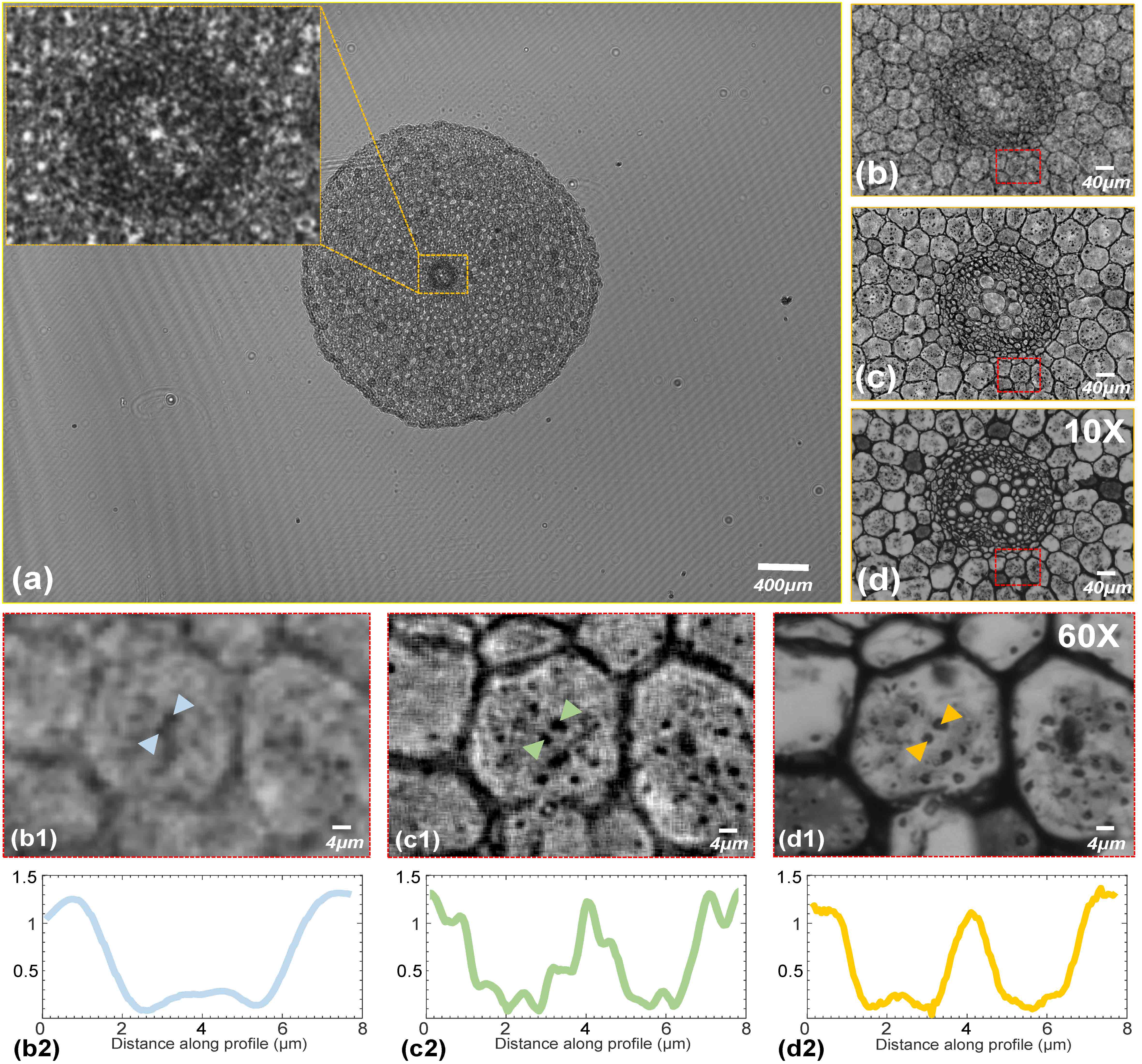}
\caption{The experimental results of a typical dicot root. (a) The full FOV low-resolution hologram. (b) The reconstructed intensity image based on the traditional method\cite{greenbaum2012maskless,gerchberg1972practical} (\(\alpha  = 0.5\), BLPEC). (c) The reconstructed intensity image based on our method. (d) The full FOV intensity image with 10X objective. (b1) and (c1) The enlargements of the red dashed boxed areas of (b) and (c) respectively. (d1) The full FOV intensity image with 60X objective corresponding to the red dashed boxed areas of (d). (b2, c2, d2) The line profiles along the respective arrow.}
\label{fig:Fig10}
\end{figure}

Another experiment was demonstrated that our method can also be used for the dense sample such as plant slice, which can be seen in Fig. \ref{fig:Fig10}. Figure \ref{fig:Fig10}(a) shows the full FOV of the typical dicot root ($\sim$$29.85$ $m{m^2}$), and the whole sample can be captured which is challenging for the traditional high-magnification lens microscope. From upper left enlarged region of the orange dashed box in raw full FOV low-resolution hologram [see the inset of Fig. \ref{fig:Fig10}(a)], we can find that the details in the typical dicot root are hard to be observed because they are submerged in the diffraction fringes. Figures \ref{fig:Fig10}(b) and \ref{fig:Fig10}(c) show the reconstructed intensity images based on the traditional method \cite{greenbaum2012maskless,gerchberg1972practical} (\(\alpha  = 0.5\), BLPEC) and our method respectively. The selected area [Fig. 10(c)] occupy only 1\% of the full FOV [Fig. 10(a)], which corresponds to the whole FOV with the 10X objective as shown in Fig. \ref{fig:Fig10}(d). From Fig. \ref{fig:Fig10}(c), it is easy to distinguish endodermis, pericycle, primary phloem, primary xylem, and parenchyma cell, which are extremely important for botanical studies. To observe the details inside the amyloplasts, we further select a small area in Fig. \ref{fig:Fig10}(c) which is the full FOV with 60X objective [Fig. \ref{fig:Fig10}(d1)]. Figures \ref{fig:Fig10}(b1) and \ref{fig:Fig10}(c1) are the local enlarge enlargements of the rectangular areas in Figs. \ref{fig:Fig10}(b) and \ref{fig:Fig10}(c) separately. As shown in the enlargements [Fig. \ref{fig:Fig10}(c1)], the grains in amyloplasts are distinguishable and the sharp improvements are noticed in the image contrast compared to the results shown in Fig. \ref{fig:Fig10} (b1). Figures. \ref{fig:Fig10}(b2)-(d2) show the line profiles along the respective arrow, and two particles in the middle cortex are easily to distinguish which is impossible in the traditional method. However, there are many horizontal and perpendicular lines in the enlargements [Fig. \ref{fig:Fig10}(c1)], and blur is obvious in Fig. \ref{fig:Fig10}(c2). The reasons are mainly that guide filter is sensitive to the smooth background, but this experimental sample is not piecewise smooth and the guided filter brings the aberration to the reconstructed results. Furthermore, the test object has a certain thickness and the diffraction patterns of the non-target objects in the vertical direction of objects on focal plane will influence imaging reconstructed results. In Supplementary Video \textcolor[rgb]{0,0,1}{2}, we show a zooming video of the full-FOV reconstruction result of a typical dicot root with our method and the traditional method \cite{greenbaum2012maskless,gerchberg1972practical} (\(\alpha  = 0.5\), BLPEC) respectively.

\section*{Conclusion}
We introduce an APLI, which can help to mitigate the artifacts and obtain super resolution images only with Z-scanning. In this paper, more than double pixel resolution of camera is successfully achieved only using intensity measurements on different the sample-to-sensor planes, and there is no extra embedding medium between the object and sensor, like the refractive index matching oil. Here we emphasize that this super-resolution technique does not require lateral displacements, wavelength changing, and illumination angles scanning. During the reconstructing process, an imaging sensor with pixel-count of 10.7 million and pixel-size of $1.67$ $\mu m$ which can acquire large FOV ($\sim$$29.85$ $m{m^2}$), is used to capture a set of out-of-focus diffracted patterns. In our experimental implementation, the samples are scanned vertically lightly and ten out-of-focus undersampling intensity images with artifacts are used to improve smallest resolved half-pitch to $0.77$ $\mu m$. Instead of the traditional fixed-step, an adaptive relaxation factor strategy has been firstly introduced into our method to suppress the over-amplification noise and retain the convergence speed under noisy conditions. Furthermore, we introduce an ALPEC method into our method which tallies with actual physical process. This method will avoid the misalignment effectively and improve the stability and robustness of the reconstruction.

We believe that our method will broadly benefit the lensfree imaging microscope and acquire higher resolution with the same amount of data comparing to the traditional reconstruction methods. In addition, our method can vastly not only remove the adverse impact of alteration in multiple systematic parameters on the reconstructed results, but also reduce the complexity of the actual operation. The results of the resolution target and botanical samples demonstrate that the proposed reconstructed method can offer a new way to make the lensfree microscope to be a competitive and promising tool for the medical care in remote areas in future.

However, some issues still deserve further consideration. Although the number of the captured images may influence the reconstructed results, the artifacts cannot be removed completely due to the trade-off between the resolution and the artifacts. Specifically, if we need to weaken the influence of the artifacts, the sample-to-sensor distance must be increased, but at the meantime the long-distance will result in failed acquisition of the high-frequency patterns because many patterns become denser suffering from more severe pixel aliasing or exceed the sensor area limits. On the other hand, in the actual situation, the resolution cannot be further improved while the number of raw images increases. We consider the reason for this phenomenon is that during the whole the process, the longitudinal error is not taken seriously. In future work, we will make effort to correct the tiny longitudinal error automatically.

\bibliographystyle{wlscirep}
\bibliography{reference}

\section*{Acknowledgements}
This work was supported by the National Natural Science Fund of China (61505081, 111574152), Final Assembly `13th Five-Year Plan' Advanced Research Project of China (30102070102), `Six Talent Peaks' project of Jiangsu Province, China (2015-DZXX-009), `333 Engineering' Research Project of Jiangsu Province, China (BRA2016407, BRA2015294), Fundamental Research Funds for the Central Universities (30917011204, 30916011322 ). C. Zuo thanks the support of the `Zijin Star' program of Nanjing University of Science and Technology.

\section*{Author contributions statement}

C.Z. proposed the idea. J.Z. and C.Z. conceived and designed the experiments. J.Z. performed the experiments and processed the resulting data. J.Z., J.S.,C.Z. and J.L. wrote the manuscript. C.Z. and Q.C. supervised the research.

\section*{Additional information}

\textbf{Competing financial interests:} The authors declare no competing financial interests.

\end{document}